# DETECTION AND INVESTIGATION OF THE PROPERTIES OF

# DARK ELECTRIC MATTER OBJECTS:

# THE FIRST RESULTS AND PROSPECTS

E. M. DROBYSHEVSKI*

*A.F.Ioffe Physico-Technical Institute, Russian Academy of Sciences*
*194021 St.Petersburg, Russia*

**Abstract.** Negative Dark Electric Matter Objects, daemons, have been detected by means of ZnS(Ag) scintillator screens. These objects are apparently relic elementary Planckian black holes. The scintillations in ZnS(Ag) are excited by electrons and nucleons ejected as the daemon captures a nucleus of Zn (or S). By catalyzing the decay of protons in the remainder of the nucleus, the daemon becomes capable of capturing a new nucleus, and so on. The time elapsed between successive captures (scintillations) is used to estimate the daemon velocity and the proton decay time (~1 μs). The flux of daemons with velocities from ~5 to ~30 km/s is ~$10^{-9}$ cm$^{-2}$s$^{-1}$ and varies with a period of 0.5 yr. This could indicate a preferred direction of the flux, which is a result of the Sun's moving relative to daemons in the Galactic disk and of their capture into helio- and geocentric orbits. Daemons crowd at the centres of the Earth, the Sun, and the Galaxy, where they catalyze proton decay, a process capable of accounting from a common standpoint a variety of observations and phenomena unexplained heretofore.

**Аннотация**. С помощью ZnS(Ag) экранов выполнено детектирование *отрицательных* Dark Electric Matter Objects – даэмонов. Это, по-видимому, реликтовые элементарные планковские черные дыры. Сцинтилляции в ZnS(Ag) возбуждаются электронами и нуклонами, испускаемыми при захвате даэмоном ядра Zn (или S). Катализируя распады протонов в остатке ядра, даэмон становится способным захватить новое ядро и т.д. По времени пролета между захватами (сцинтилляциями) оценивается скорость даэмона и время распада протона (~1 мкс). Поток даэмонов со скоростями от ~5 до ~30 км/с достигает ~$10^{-9}$ см$^{-2}$с$^{-1}$ и меняется с периодом 0.5 года. Это может указывать на выделенное направление вследствие движения Солнца относительно даэмонов диска Галактики и говорит об их захвате на гелио- и геоцентрические орбиты. Даэмоны концентрируются к центрам Земли, Солнца и Галактики, где они катализируют распад протонов, - процесс, способный с единой точки зрения объяснить множество непонятных ранее фактов и явлений.

## 1. Introduction. The Issue of the Dark Matter

It was believed about five years ago that 95-98% of the mass contained in the galaxies and the Universe as a whole is made up of a non-barionic dark matter (DM). This issue appeared in the 1960s in connection with the discovery of a non-Keplerian motion of stars in galaxies and of galaxies in clusters. The recent establishment of an accelerating expansion of distant objects and of the existence of a nonclamping "quintessence" (dark energy and the like) reduced the DM fraction in the galaxies down to ~80%. Nevertheless, its nature still remains unknown, and the issue has not lost its acuteness. The DM density in the galactic halo is ~0.3÷1 GeV·cm$^{-3}$.

Many hypotheses on the nature of the DM objects (massive neutrinos, brown dwarfs, black holes and similar MACHOs) have either been rejected or not yet confirmed. Most of the efforts are focused at present on a search for the elementary particles predicted by theory beyond the Standard Model, more specifically, various massive (~100 GeV) particles which only weakly interact with ordinary matter, WIMPs, among them axions, neutralinos, etc. This search is pursued under low background conditions, at large depths underground by numerous teams and for many years, assisted by the most sophisticated instrumentation [1-6]. No conclusive results have thus far, however, been produced. The only group that claims to have obtained some evidence for a definite result is the DAMA collaboration, which in the seven years of exposure of a 100-kg NaI(Tl) detector has revealed indications of a ~7% yearly modulation of weak (≤6 keV) signals at a ~6σ (!) level [7]. This variation is assumed to be related to the Earth's orbital motion superposed on the motion of the Sun relative to the DM objects of the galactic halo. Their flux onto the Earth should pass through a maximum on June 2, and through a minimum, on December 2 (the DAMA extrema occur one week earlier). At present, this variation is the only experimental evidence for the existence of WIMPs. But even this variation is real, it remains unproven that it is the WIMPs that are responsible for it [6]. Therefore detection of some more indications of the reality of the WIMPs would be highly desirable. The conclusions presented by DAMA still have not received independent support and are disputed by other groups (see discussion in Ref. 7). The lack of hard evidence and of promising approaches in other areas of this search may even prompt a suggestion that the DM objects have properties making their detection impossible altogether [8].

* E-mail: emdrob@mail.ioffe.ru

## 2. Our Conception of the Situation

We started from a simple hypothesis [9] that since our Universe started its evolution from Planckian scales, it appears only natural to assume that a large fraction of its mass still remains bound in objects of this scale. These are black holes with the smallest possible mass; a smaller mass could not stay hidden under the gravitational radius $r_g = 2GM/c^2$, because its Compton wavelength $\lambda_c = \hbar/Mc$ would be in excess of $r_g$. Whence $M \approx 2\times10^{-5}$ g, and $r_g \approx 3\times10^{-33}$ cm. Based on considerations of a fairly general nature, one may conclude that such objects should be stable and do not evaporate any more, so that they could be remnants from evaporation of more massive black holes [10], as well as that they may carry an electric charge of up to $Ze \sim G^{1/2}M \sim 10e$. Such particles (Planckeons, quantum maximons, pyrgons, Newtorites etc.) were discussed on many occasions, including the possibility of their being cold DM objects [10-14], although their existence in this form comes in a certain conflict with the well-known inflationary scenarios [15]. Besides, the possibility of their detection aroused skepticism, because at velocities ~100 km/s even charged particles are incapable of producing a scintillation [11]. The extremely small flux of such massive particles from the galactic halo (~$10^{-12}$ cm$^{-2}$s$^{-1}$), a level at which only one particle would cross a square ~2 m on a side in a year likewise did not inspire optimism in potential experimenters.

We succeeded in showing that the situation with detection of charged Planckian objects is far from being so hopeless [having recalled the concept expressed by Thales of Miletus (624-545 BC) that "παντα δαιμονων πληρη" («all things are full of gods»), we called them DArk Electric Matter Objects, daemons].

First, we took into account that, by estimates of Bahcall *et al* [16], DM objects of the galactic disk could exist in a higher concentration than those present in the halo and could also move with a small velocity dispersion (~4-30 km/s) compared to the 200-300 km/s range expected for those of the galactic halo. If these are daemons carrying an electric charge, they should be rather efficiently slowed down in transit through the Sun and end up in becoming captured by it into heliocentric orbits [17]. A combined action of the Sun and of the Earth would transfer some of them into strongly elongated, Earth-crossing heliocentric orbits (SEECHOs) with a perihelion outside the Sun [18]. The perturbing action of the Earth would subsequently transfer part of the daemons from these into near-Earth, almost-circular heliocentric orbits (NEACHOs). Our estimates (which presently seem to have been too optimistic) showed that the flux of SEECHO daemons at the Earth's surface could be as high as ~$3\times10^{-7}$ cm$^{-2}$s$^{-1}$, which should exceed by four to five orders of magnitude that of daemons from the galactic halo. This estimate made detection of low-velocity ($V \sim$ 10-50 km/s) daemons, members of the Solar system, a promising experiment [18].

Second, *negative* daemons interact strongly with matter at the nuclear level [19]. Indeed, like muons, they are capable of *catalyzing the fusion of light nuclei*. Moreover, for $Z = 10$ their ground level lies within the captured deuteron, and for a nucleus charge $Z_n \geq 24/Z$, even inside a proton in a nucleus. The binding energy with a nucleus can be estimated as $W = 1.8ZZ_nA^{-1/3}$ MeV [20], which for the Fe or Zn nucleus is in excess of 100 MeV. Therefore, the capture of a nucleus is accompanied by (*i*) emission of Auger and conversion (including refilling) electrons, followed by (*ii*) evaporation of a part of the nucleons and their clusters from the nucleus. While staying in the remainder of the nucleus inside a proton, the daemon will disintegrate it [19]. There is a close analogy with the monopole catalysis of the fast ($\sim 10^{-6\pm2}$ s) proton decay pointed out by Rubakov [21]. The fields generated in the immediate vicinity of a daemon are by definition the strongest in Nature. Therefore, while we do not know details of the processes involved (there is still no theory of quantum gravitation), one may judiciously assume that a daemon-containing proton should decay somehow in an extremely short time. Our estimates of the time needed for a daemon to recover its catalytic capability, which were made under the assumption of the solar energetics being based to a considerable extent on the daemon-assisted catalysis of proton fusion, yielded $\sim 10^{-7}$-$10^{-6}$ s for the proton decay time [22] (we know now that our approach was not fully correct; indeed, proton decay itself can be a more efficient source of the solar energy - see Sec. 7 below - but nevertheless, these estimates formed a basis for the development of a real operating detector).

The above reasoning yielded a few more conclusions of practical significance:
(1) The scintillation technique should be an efficient tool for the detection of daemons;
(2) In view of the fact that the capture of heavy nuclei by a daemon is accompanied by ejection of many particles, inorganic scintillators consisting of heavy elements (ZnS, CsI etc.) would be preferable;
(3) To carry out detection experiments in standard (not underground) conditions against the background produced by cosmic rays (their flux $\sim 10^{-2}$ cm$^{-2}$s$^{-1}$), it is desirable to use thin-layer ($\sim$10 μm) scintillators. In crossing such a thin layer, the protons ejected out of the nucleus loose $\sim$1 MeV of energy, whereas light relativistic particles of the cosmic rays (mesons etc.) release <10 keV in the layer.

## 3. Technical Realization of the Detector and the First Results

Our primary goal was not to perform high-precision quantitative measurements but rather to detect at least any indication of the existence of daemons. We had to "hook" onto something.

Our first attempt consisted in trying to detect SEECHO daemons as they cross a block of Li or Be metal and catalyze the fusion of these nuclei. While for a number of reasons that now appear obvious these experiments were unsuccessful [23], they served as a basis for formulation of the most essential aspects of the issue of daemon interaction with matter, particularly in what concerns the fundamental role played by the proton decay catalysis, aspects that appear presently fairly obvious. This analysis culminated in construction of a simple detector made up of four identical modules [24].

Each module (Fig. 1) consists of two transparent, polystyrene plates, 4 mm thick and 0.5×0.5 m$^2$ in area, which are light-isolated from one another and spaced 7 cm apart. They are arranged in the middle of a cubic tinned-iron case (0.3 mm Fe coated on both sides by 2 μm of tin), 51 cm on a side, whose top side is covered by a sheet of black paper. Each plate is viewed on its side by a FEU-167 PM tube (photocathode diameter 100 mm) from a distance of 22 cm. The bottom side of each plate is coated by a ~3.5 mg·cm$^{-2}$ layer of ZnS(Ag) scintillator powder. To enhance the difference between the signals caused by the objects we are looking in crossing the plates up- and downward, we purposefully introduced an up/down asymmetry into the detector design.

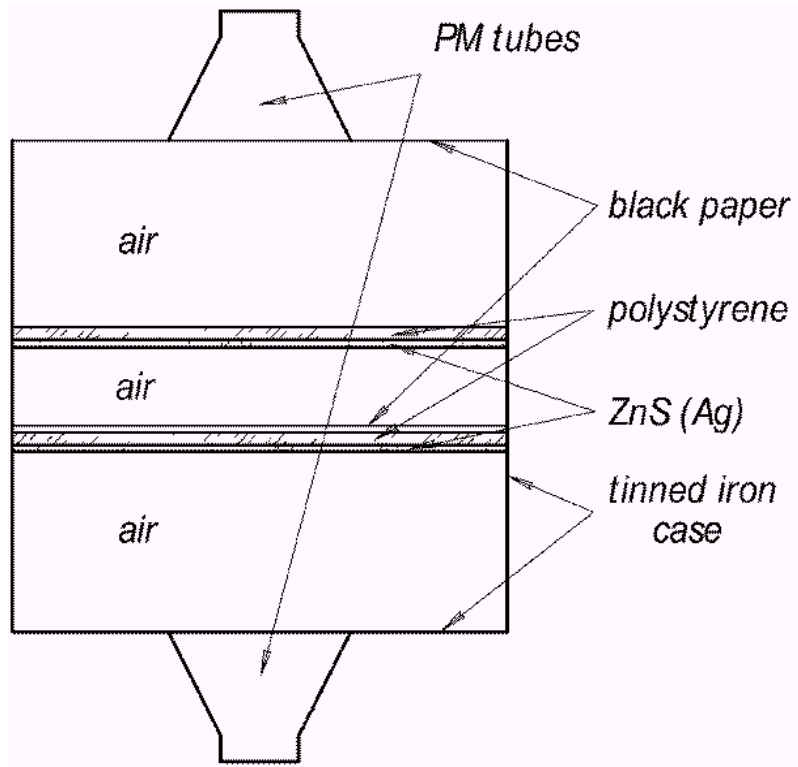

Fig. 1. Schematic of a module for detecting the time-shifted scintillations.

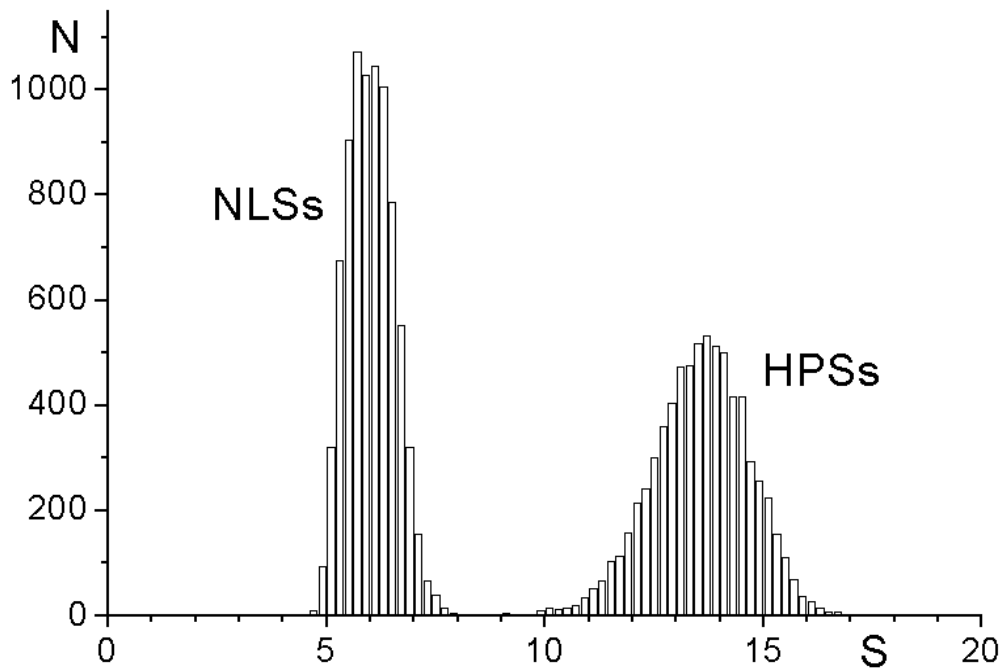

Fig. 2. Typical distribution of the NLSs and HPSs versus the scintillation form-factor *S*, the signal-amplitude normalized area swept by the oscilloscopic trace of the scintillation. These distributions can be approximated by Gaussians. The NLS and HPS distributions are well distinctive by their *S* values. The "narrow" HPSs and "wide" HPSs lie, respectively, at the left and right-hand parts of the HPS Gaussian.

Signals from the top and bottom PM tubes were fed into a digital dual-trace oscilloscope and were observed within the ±100 µs interval from the trigger pulse produced by the top PM tube. If the bottom trace had also a signal, such event was saved in computer. Only those events that produced one signal on each trace, with the beginning of the bottom signal shifted by $|\Delta t| > 0.5$ µs with respect to that of the top signal, were processed, which means that the cosmic rays were excluded from consideration (they were used, however, to compare the sensitivities of the top and bottom channels).

ZnS(Ag) is not a spectrometric scintillator. But it consists of components with a large atomic weight and has a high light yield (~28% of the energy released in it [25]). Besides, preliminary experiments which included a study of intrinsic PM tube noise, detection of cosmic rays (simultaneous signals from the top and bottom PM tubes, sometimes in more than one module), and calibration using ~200 keV X-rays and 5.5-MeV α-particles from a $^{238}$Pu source showed that the shape of signals from ZnS(Ag) depends strongly on the actual kind of the radiation that produced them. A scintillation signal due to a heavy nonrelativistic particle (α-particle) has a smooth maximum 2.5-3 µs from its beginning; it is referred to here as Heavy Particle Scintillation (HPS). Signals originating from cosmic rays are characterized by a short rise time (<1 µs), and their trailing edges are fully determined by the PM tube anode load resistance. In this respect they do not differ from the PM intrinsic noise, which accounts for our referring to them as Noise-Like Scintillations (NLSs). Figure 2 presents typical distributions of the NLS and HPS background events plotted vs. the formfactor $S$, i.e., the area of the scintillation oscillogram normalized to its amplitude [26]. The distributions are seen to be well separated from one another, and each of them can be well fitted by a Gaussian.

To reliably isolate the HPSs peaking near 2.5 µs and suppress the NLSs, we used in the initial experiments a 4.5 kΩ resistor for the PM tube anode load, and the output signal was fed to the oscilloscope through an inductance of 4.69 mH and a cable of total capacitance 550 pF (the circuit oscillation period was 10 µs, with the maximum reached in 2.5 µs). As a result, the HPS amplitude remained almost unchanged, whereas the amplitude of the NLSs decreased three to five times. The number of recorded NLS signals also decreased accordingly.

Our approach consisted in looking for events occurring on the upper and lower oscilloscopic traces which would be correlated in $\Delta t$. This approach was based on the following two, fairly obvious postulates:

*Postulate A*. The distribution of the number of events $N(\Delta t)$ should not differ from a constant if it is caused by purely accidental, unrelated background events in both ZnS(Ag) screens. This distribution can differ from $N = const$ only if there is a contribution of the daemon component (if we exclude the manifestation of conventional cosmic rays with $\Delta t \approx 0$);

*Postulate B*. Any passive factor capable of distorting somehow $N(\Delta t)$ should suppress the contribution of the daemon component to $N(\Delta t)$. This should result in $N(\Delta t) \rightarrow const$, with very rare exceptions where only a part of the contribution due to the daemon component is mainly suppressed, in which case its other parts may become manifest more clearly.

Figure 3 presents a $N(\Delta t)$ distribution for 413 events recorded in March 2000 during 700 h ≈ $2.5 \times 10^6$ s [24]. According to the $\chi^2$ criterion, the $N(\Delta t)$ distribution cannot be approximated by a constant at a confidence level 1 - $\alpha$ = 99.8%. If we select only the HPS events recorded on the upper trace, the number of events drops to 212. Nevertheless, the statistical significance of $N(\Delta t)$ ≠ *const* remains almost the same (1 - $\alpha$ = 99.85%). In both cases, a peak in the 20 < $\Delta t$ < 40-µs bin stands out. In the first case it contains 62 events, which exceeds the mean level of 41.3 event/bin by ~2.6√62 = 2.6$\sigma$, while in the second it retains 39 events but still is above the mean level of 21.2 event/bin by 17.8 events, i.e., by ~2.85√39 = 2.85$\sigma$. Thus, the statistical significance of this maximum is 99.5%.

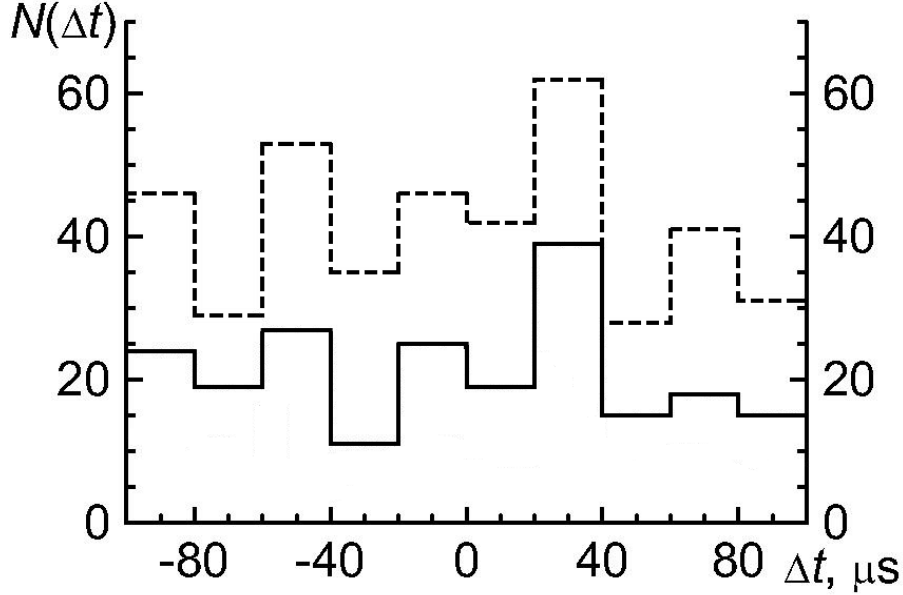

Fig. 3. (---) Distribution $N(\Delta t)$ of pair scintillation events on their time shift (relative to the upper channel events). (—) Similar distribution which takes into account only HPS (heavy-particle scintillation) type events at the upper channel.

Assuming the maximum near $\Delta t \approx +30$ μs to be due to a downward flux of some particles, the flux calculated for a total detector area of 1 m$^2$ becomes $f_\oplus \approx 0.7 \times 10^{-9}$ cm$^{-2}$s$^{-1}$. Including the particle flux outside this maximum and from below (for $\Delta t < 0$), as well as the reduced light collection efficiency from the peripheral zones of the scintillation screens, can increase $f_\oplus$ by approximately an order of magnitude.

The detection of a statistically significant peak at $\Delta t \approx +30$ μs instilled optimism while at the same time raised several questions:
(1) Could not the results of the measurements and, in particular, this peak be due to errors in designing the experiment, functioning of the instrumentation, or to some other artifacts?
(2) Why is the velocity derived from the 7-cm base between the ZnS(Ag) screens and the time interval $20 < \Delta t < 40$ μs so small ($V \approx 3.5$-$1.8$ km/s)? Taking into account the possible trajectory inclinations within the limits defined by the case walls can at most double these values to ~4-7 km/s.

## 4. Further Experiments: Gaining a Deeper Insight into the Properties of Daemons

To answer these questions, additional experiments involving variation of the parameters of the detectors and of the measuring equipment were undertaken. During a whole year of these tests, no instrumentation malfunctions were revealed.

Our concern was aroused initially by a lack of reproducibility in the results. Raised in the traditions of conventional laboratory work, we got used to the idea that a controlled variation of any parameter of an experimental setup should lead to some consequence, and returning this parameter to its original value should produce the initial reading. The situation with this experiment was radically different. Because acquisition of the desired statistics with four modules takes up at the very least a month, check measurements stretch naturally to many months. Therefore progress in understanding the properties of the detector and of the processes involved in the interaction of daemons with it was not as fast as we would like it. To speed it up, a second four-module detector was put in operation by the mid-2000, and in order not to lose weak signals, the inductances were removed from the PM tube anode circuits by the end of March 2001. We now used a standard resistance anode load (30 kΩ) for the cable capacitance of 150 pF (in the hope that the action of inductance could always be imitated numerically); true, the oscilloscope triggering level had to be increased from ~2.5 to 8-10 mV. Nevertheless, in the course of a year-long process of trial and error modifications and of comparing the results of our experiments with ideas concerning the evolution of daemons in space subject to the laws of celestial mechanics, we finally came to the understanding of some new aspects of the problem:

(1) The lack of reproducibility from one month to another is caused primarily by a strong seasonal variation of the daemon flux, which implies that our starting hypothesis, namely, the simplest assumption of their flux onto the Sun being practically isotropic is wrong. This means that the Sun moves with respect to the daemon background in the galactic disk with a velocity comparable to the dispersion of their velocities. This agrees with the estimate of Bahcall *et al* [16] of the dispersion in the velocities of possible disk DM objects of 4-30 km/s, a figure comparable to the Sun's velocity relative to the nearest stars of ~20 km/s. In view of the capture of daemons into SEECHOs as a result of their slowing down in transit through the Sun, the direction we detect them from suggests the variation of their flux, generally speaking, with a half-year period, which corresponds to the Earth's transit through a shadow and an "antishadow". The first is created by the Sun in the flow of the incoming daemons from the galactic disk population. The "antishadow" forms as the shadow objects slowed down (and captured) by the Sun fall back on and cross it.

(2) The correct value of the base distance in the detector that should be employed in calculations of the daemon propagation velocity became clear. The fact is that we chose the 7-cm gap between the scintillation screens in an attempt to detect the passage of the SEECHO objects with a velocity ~50 km/s (or somewhat smaller) for a daemon-stimulated proton-decay time $\Delta\tau_{ex} \sim 10^{-7}$ s. In this case, after the capture by a daemon in ZnS(Ag) of a Zn nucleus with an energy release $W \approx 130$ MeV the nucleus undergoes de-excitation by evaporating ~12 nucleons (including 5-6 protons). To be capable of capturing another nucleus when entering the bottom scintillator, the daemon has to disintegrate in the remainder of the nucleus ~15-16 protons, after which the net charge of the daemon and the nucleus remainder will become negative. For $\Delta\tau_{ex} \sim 10^{-7}$ s, the protons will decay in ~1.5 μs, and it is this figure that requires a gap of ~7 cm for the velocity ≤50 km/s. As we have seen above, however, this yields for $\Delta t \approx 30$ μs the unreasonably small value $V \leq 4$-7 km/s.

The 30-μs peak yields a reasonable estimate for the velocity if one takes the 29-cm distance between the top ZnS(Ag) screen and the bottom tinned-iron lid as the base.

It appears appropriate to suggest that the decay of ~15-16 protons in the remainder of the Zn nucleus come to an end already after the downward moving daemon has crossed the bottom ZnS(Ag) screen, so that the capture of a new heavy nucleus with emission of energetic electrons and nucleons can take place in the bottom lid of the case only. It is these particles, primarily electrons, that reach the bottom ZnS(Ag) layer and excite scintillations in it. Taking into account the trajectory inclination, the distance of 29 cm yields $V \approx 10$-15 km/s. This velocity corresponds to the fall of objects from near-Earth, almost circular heliocentric orbits (NEACHOs), into which they should be transferred by Earth perturbations from the SEECHOs. Recalling the reasoning in the above paragraph, we obtain immediately an upper bound on the daemon-stimulated proton decay time $\Delta\tau_{ex} \leq 30$ μs/(15-16) = 2-1.8 μs.

(3) We understand now better some details in the interaction of daemons with matter and, in particular, with the detector components. The interaction cross section of a very fast particle with a nucleus is determined by the geometric cross section of the latter, $\sigma_0 \approx \pi R_0^2 A^{2/3}$, where $R_0 = 1.2 \times 10^{-13}$ cm, and $A$ is the nuclear atomic weight. The cross section for slow negative particles with $Z_{eff}$ increases because of their attraction and displacement of the nucleus [23]

$$\sigma = \sigma_o (Z_{eff} Z_n e^2 / R_0 A^{1/3}) \times (2/A m_p V^2), \qquad (1)$$

Table 1. Parameters of the detector components and quantities characterizing their interactions with daemons at three typical velocities ($V = 5$, 15, and 45 km/s). $l$ is the component size and $t_l$ is time of its traversing by daemon; $\lambda$ is the mean free path needed for a daemon with $Z_{eff} = 1$ (the most probable case) to capture a nucleus, $\tau_\lambda$ is the corresponding time (to calculate $\lambda$ and $\tau_\lambda$, the cross-section defined by Eq.1 is used). $K = \lambda \, Z_{eff}/l$ is the Knudsen number.

| | $Z_n$ | A | $\rho$ [g/cm³] | $l$ [cm] | | | | $t_l$ [μs] | | | $\lambda \cdot Z_{eff}$ [cm] | | | $\tau_\lambda \cdot Z_{eff}$ [ns] | | | $K = \lambda\, Z_{eff}/l$ | | |
|---|---|---|---|---|---|---|---|---|---|---|---|---|---|---|---|---|---|---|---|
| V [km/s] | | | | | 5 | 15 | 45 | 5 | 15 | 45 | 5 | 15 | 45 | 5 | 15 | 45 | 5 | 15 | 45 |
| H | 1 | 1 | (0.082) | | | | | | | | 4.9·10⁻⁵ | 4.4·10⁻⁴ | 4.0·10⁻³ | 0.1 | 0.29 | 0.89 | | | |
| C | 6 | 12 | (0.978) | | | | | | | | 5.2·10⁻⁵ | 4.7·10⁻⁴ | 4.2·10⁻³ | 0.1 | 0.31 | 0.93 | | | |
| poly-styrene | CH | | 1.06 | 0.4 | 0.8 | | 0.09 | | 0.27 | | 2.5·10⁻⁵ | 2.3·10⁻⁴ | 2.0·10⁻³ | 0.05 | 0.15 | 0.44 | 6.3·10⁻⁵ | 5.6·10⁻⁴ | 5·10⁻³ |
| N | 7 | 14 | (0.00091) | | | | | | | | 5.1·10⁻² | 4.6·10⁻¹ | 4.1 | 102 | 307 | 910 | | | |
| O | 8 | 16 | (0.00028) | | | | | | | | 0.18 | 1.6 | 14.5 | 360 | 1070 | 3200 | | | |
| air | | | 0.0012 | 7 | 14 | 4.7 | 1.6 | | | | 0.04 | 0.36 | 3.2 | 80 | 240 | 710 | 5.7·10⁻³ | 5.3·10⁻² | 0.45 |
| | | | | 22 | 44 | 14.7 | 4.9 | | | | | | | | | | 1.8·10⁻³ | 1.6·10⁻² | 0.15 |
| S | 16 | 32 | (1.35) | | | | | | | | 5.8·10⁻⁵ | 5.2·10⁻⁴ | 4.7·10⁻³ | 0.12 | 0.35 | 1.04 | | | |
| Zn | 30 | 65 | (2.74) | | | | | | | | 5.1·10⁻⁵ | 4.5·10⁻⁴ | 4.1·10⁻³ | 0.10 | 0.30 | 0.91 | | | |
| ZnS(Ag) | | | 4.09 | 10⁻³ | 2·10⁻³ | 0.7·10⁻³ | 2.2·10⁻⁴ | | | | 2.7·10⁻⁵ | 2.4·10⁻⁴ | 2.2·10⁻³ | 0.054 | 0.16 | 0.49 | 2.7·10⁻² | 0.24 | 2.17 |
| Fe | 26 | 56 | 7.9 | 0.03 | 0.6 | 0.02 | 0.007 | | | | 1.6·10⁻⁵ | 1.4·10⁻⁴ | 1.3·10⁻³ | 0.032 | 0.093 | 0.29 | 5.3·10⁻⁴ | 4.7·10⁻³ | 4.3·10⁻² |
| Sn | 50 | 119 | 7.3 | 2·10⁻⁴ | 4·10⁻⁴ | 1.3·10⁻⁴ | 4.4·10⁻⁵ | | | | 3.1·10⁻⁵ | 2.8·10⁻⁴ | 2.5·10⁻³ | 0.062 | 0.19 | 0.56 | 0.16 | 1.4 | 12.5 |

where $m_p$ is the proton mass. We disregard here the excitation and possible deformation of the nucleus, which would increase $\sigma$.

Table 1 (Ref. 26) gives an idea of the mean free paths $\lambda$ and of the corresponding times $\tau_\lambda$ between the collisions of a negative supermassive particle with $Z = Z_{eff}$ with nuclei in the components of our detector (polystyrene, air, ZnS(Ag), iron, tin) calculated for three values of the velocity $V = 5$, 10, and 45 km/s. The table presents also the dimensions $l$ of the detector components and the corresponding transit times $t_l$. Interestingly, low-velocity particles ($V \leq 30$ km/s) have a very high probability for colliding in a 10-µm layer with a Zn or S nucleus and capturing it with excitation. Particles with a higher velocity ($V \geq 40$ km/s) are capable of crossing the 10-µm scintillators undetected, because they do not collide with nuclei. On the other hand, in passing through tinned iron of the case, a daemon with $Z_{eff} \geq 1$ will inevitably interact with an iron nucleus. By contrast, slow objects with $V \leq 10$-15 km/s will first be "poisoned" by a Sn nucleus in the thin tin layer. It is this shielding of the scintillating screens by the side walls of the case which cut out a solid angle of ~2 ster that can account for the narrowness of the peak in the interval $20 < \Delta t < 40$ µs. Capture of a Sn nucleus in the bottom case lid by upward moving daemons with $V \sim 10$-15 km/s can explain also the absence of a peak at $-40 < \Delta t < -20$ µs. The energy released in the capture, $W \approx 183$ MeV, is high enough to evaporate ~18 nucleons, including ~9 protons. There will be ~41 protons still left in the remainder of the nucleus. Recalling that in order for a scintillation to be produced in the top ZnS(Ag) screen by capture of a nucleus, the charge of the latter should be $Z_{eff} \leq 9$, this yields an estimate $\Delta\tau_{ex} \geq 30$ µs/(41 - 9) $\approx 0.9$ µs.

## 5. Semi-Annual Variation of the Daemon Flux

By summer 2001 it became clear that the main reason for the lack of reproducibility in our observations is, among other factors, a seasonal variation of the daemon flux, and it was decided to avoid, wherever possible, any modifications in the four operating modules, however desirable they would seem, in order for these modules with Ohmic load in their PMT anode circuits to provide a uniform series of observations during several years (May and the beginning of June 2001 were spent to transfer the detectors to rooms with controlled temperature, first to the basement, where we were confronted by the problem of radon, and afterwards, to a small room with an air conditioner). The results of all measurements were now also processed by the same technique taking into account the shape of the HPSs. In this case, we abandoned analysis of fine features in the $N(\Delta t)$ distribution. In accordance with *Postulate A* (see Sec. 3), we chose as the main criterion characterizing the existence of a daemon flux the confidence level 1 - $\alpha$ that $N(\Delta t) \neq const$ within $-100 < \Delta t < 100$ µs. The calculation of 1 - $\alpha$ was carried out with the use of the $\chi^2$ criterion.

A few words would be appropriate here for better understanding of what follows. The value of $\chi^2$ for a statistically random distribution $N(\Delta t)$ is in the limit always equal to the number of degrees of freedom $\varphi = n - 1$, where $n$ is the number of bins. In our case, $-100 < \Delta t < 100$ µs, which means that for a bin width of 20 µs we have $n = 10$, so that $\chi_{lim}^2 = 9$. For $\chi^2 = \varphi = 9$, this corresponds, on the average, to 1 - $\alpha = 0.56$. Therefore, in the absence of factors capable of biasing the distribution from the hypothesis under test (in our case, this is $N(\Delta t) = const$), the values of 1 – $\alpha$ for different samples of $N(\Delta t)$ should lie in the vicinity of 0.56, but, as a rule, within somewhat uncertain limits $0.1 < 1 - \alpha < 0.9$ (see e.g. [27]). That 1 - $\alpha$ is larger than 0.9 may be considered as a compelling argument for the deviation of $N(\Delta t)$ from *const* being real.

Month-by-month cosmic ray statistics was used to determine the relative sensitivity of the scintillation detection system-PM tube-amplifier of the top and bottom channels of each module. After this, the computer calculated a table whose cells corresponded a given number $N_{mod}$ of events that were recorded in each module during a specified time interval (~ a month). The events selected were those whose oscilloscopic amplitude was in excess of their lowest levels in the top ($U_{1min}$) and bottom ($U_{2min}$) channels, with the fixed ratio $\omega_0 = U_{1min}/U_{2min}$ being the second parameter determining the cell of the table. Such an approach, where only events with a large enough signal amplitude are selected out of their total large number for further analysis, makes scrupulous calibration of the system superfluous; indeed, the only thing left in this case is to check that the daily number of events in each module, which are mainly background signals, is about constant. The number of events recorded in each module in a month varied in actual practice from 300 to 500. The probabilities 1 - $\alpha$ for the distributions $N(\Delta t)$, $N_w(\Delta t)$, $N_n(\Delta t)$, and $N_w(\Delta t) + N_n(-\Delta t)$ were also entered into each cell. An analysis of this table shows cells with maximum values of 1 - $\alpha$. For $N(\Delta t)$, they usually correspond to $N_{mod}$ = 80-90 and $\omega_0$ = 4-5.

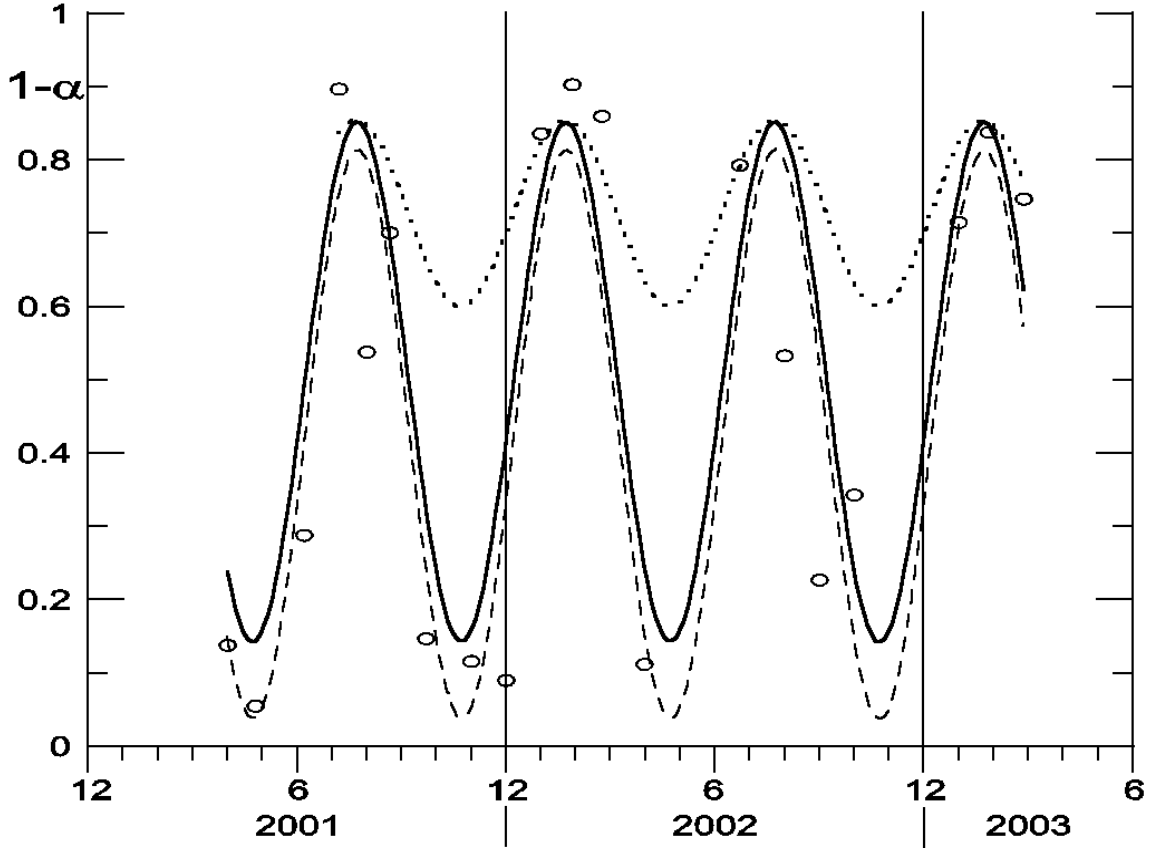

Fig. 4. Seasonal variation of 1 - $\alpha$, the extent to which the $N(\Delta t)$ distribution (at -100 < $\Delta t$ < 100 μs) deviates from the constant level produced by background events. (- - - -) Weights of all the points are equal, the correlation coefficient of the sine curve ($P$ = 0.5 yr) with the points is $r$ = 0.86, its C.L. > 99.9%.
(———) Weights of the points are proportional to 1 -$\alpha$; $r$ = 0.75, its C.L. = 98.7%.
(········) Points with 1 - $\alpha$ > 0.5 are considered only; their weights are equal to 1 - 2$\alpha$; $r$ = 0.36, its C.L. ≈ 50%.

The results revealed with this approach turned out to be rather interesting. Figure 4 presents experimental values of 1 - $\alpha$ obtained for $N_{mod}$ = 90 and $\omega_0$ = 4 during two years of operation (see also Ref. 28).

We had yielded earlier to the temptation and drew a sine curve with $P$ = 0.5 yr through all points obtained in different months, assuming them to be equally significant, irrespective of their value of 1 - $\alpha$. We could not find in textbooks on mathematical statistics clear-cut recommendations on a correct approach to follow in such cases. This time we addressed this problem more cautiously and considered two additional versions by assigning different weights to the points in Fig. 4, namely, (*i*) a weight of 1 - $\alpha$ or (*ii*) a weight of 1 - 2$\alpha$ for 1 - $\alpha$ > 0.5, and 0 for 1 - $\alpha$ < 0.5, respectively.

Then the totality of the experimental values of 1 - $\alpha$ obtained in two years, despite their considerable scatter, can be fitted by sine curves with a period of 0.5 year, with correlation coefficients depending on the accepted weights of the points. The statistical significance of the correlations also depends on the weights (the numerical values are specified in the caption to Fig. 4).

The deviation of $N(\Delta t)$ from a constant level reaches a maximum somewhere in February and August, which is roughly in agreement with the direction of the Sun's motion relative to the nearest stellar population. Thus, our ideas concerning the existence of a shadow and antishadow in the daemon flux impinging on the Sun, and the capture of daemons in this region into the SEECHOs and, subsequently, into the NEACHOs, find a support. It is clear, however, that further improvement of statistics is needed in order to draw more reliable conclusions.

## 6. Taking into Account the HPS Shape, and Detection of a Low-Velocity Population in Geocentric Earth-Surface-Crossing Orbits (GESCOs)

In Sec. 3 (see Fig. 3), the difference in shape between the HPSs and NLSs was already used in interpretation of the experimental data and isolation of physically more significant information.

It appears natural to assume that the shape itself of an HPS initiated by a daemon should depend on the direction of its propagation, up- or downward, through the detector. Note that an HPS excited in the capture of a nucleus by a daemon consists, generally speaking, of two parts, namely, of an NLS generated by Auger and other (conversion, refilling, etc.) electrons, and an HPS itself produced by the nucleons emitted from the nucleus. Therefore, if a daemon is propagating, for instance, upward with a fairly high velocity, part of the nucleons will be emitted from the nucleus, which becomes excited in the capture in ZnS(Ag), already in polystyrene and, thus, will not be involved in creation of a scintillation, with the result that the scintillation will become shorter, and the formfactor $S$ will decrease. In view of the poor transparency of the ZnS(Ag) powder to the intrinsic radiation [25], it may be expected that the situation will reverse with decreasing daemon velocity, so that the fraction of the NLS component for the daemon propagating, for instance, upward, will decrease, and that of the HPS, increase, because of redistribution of the zones in the scintillator layer where the electrons and nucleons are primarily emitted. A downward moving object will produce the reverse situation [26, 28].

Thus, one could expect that the $N_w(\Delta t)$ and $N_n(\Delta t)$ distributions constructed separately for the "broad" and "narrow" HPSs corresponding to the right and left wings of an HPS Gaussian, which is illustrated by Fig. 2, should be to some extent in anticorrelation [26].

As seen from Fig. 5 presenting the $N_n(\Delta t)$, $N_w(\Delta t)$, $N(\Delta t) = N_w + N_n$ distributions and the "inverted" distribution $N_n(\Delta t) + N_w(-\Delta t)$ for April 2001, this expectation is to a considerable extent confirmed. The inverted distribution can be approximated by a straight line $a\Delta t + b$, where $a < 0$. Moreover, the existence in $N_w$ and $N_n$ of clearly pronounced maxima at $|\Delta t| > 40$ μs suggests the appearance of a population of daemons propagating with a velocity $V < 6-7$ km/s both downward (the peak at $\Delta t > 40$ μs for $N_n$) and upward (the peak at $\Delta t < -40$ μs for $N_w$), an observation of particular interest. We can draw the following conclusions:
(1) Daemons from the NEACHOs are captured into geocentric, Earth-surface-crossing orbits (GESCOs).
(2) Because the peak at $\Delta t \approx 30$ μs (Fig. 3) shifts in a month to $|\Delta t| > 40$ μs (Fig. 5) and shifts in May beyond $|\Delta t| = 100$ μs to become practically unobservable [26], one can readily estimate the time of the slowing-down and its force acting on the daemon as it crosses the Earth. Assuming the slowing-down time to be ~2 months, and the orbital motion period in GESCO of ~6000 s, one comes for the slowing-down force to ~$10^{-10}$ N [29].
(3) The appearance of a maximum at $\Delta t < -40$ μs indicates that an upward-moving daemon is

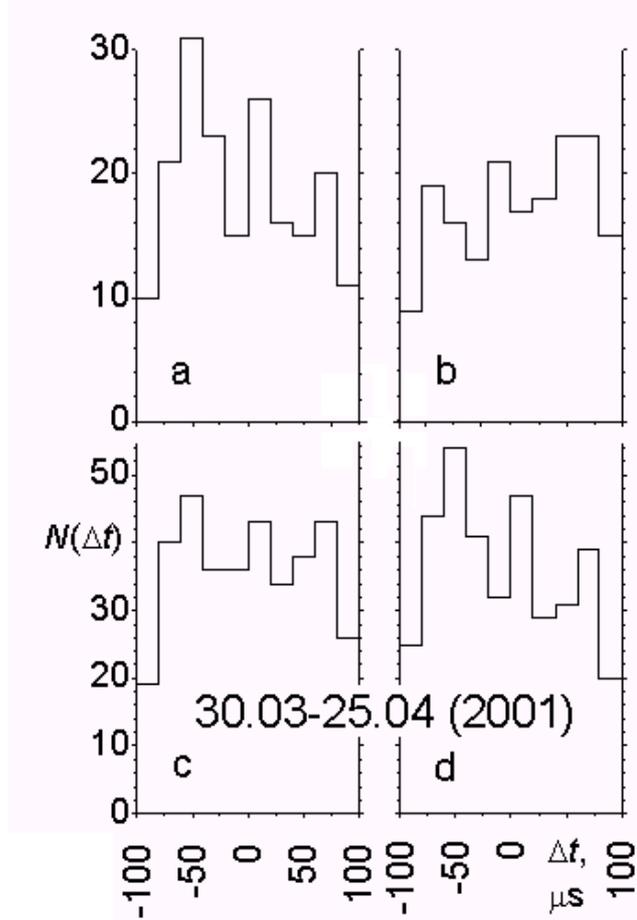

Fig. 5. Statistics for shifted scintillations of upper and lower scintillators. Only HPS events at the upper scintillator are taken into account. (a) the "wide" HPS distribution $N_w(\Delta t)$; a maximum at -80 < $\Delta t$ < -40 µs demonstrates an existence of the daemon flux going through the Earth upwards; (b) the "narrow" HPS distribution $N_n(\Delta t)$ (for the definition of $N_w(\Delta t)$ and $N_n(\Delta t)$ see Fig. 2); (c) the combined distribution $N(\Delta t) = N_w + N_n$; (d) the "inverted" combined distribution $N_n(\Delta t) + N_w(-\Delta t)$, which shows the dependence of the HPS width on the up/down direction of the deamon motion.

capable of digesting during this time a Sn nucleus captured in the lower box lid. Repeating the estimates made at the end of Sec. 4 but now for $\Delta t$ < -40 µs, we obtain $\Delta\tau_{ex} \leq 1.3$ µs, which, in view of the above estimates, yields $0.9 \leq \Delta\tau_{ex} \leq 1.3$ µs for the *daemon-stimulated proton decay time*.

It is appropriate to make here a few methodological remarks concerning the processing of experimental data. As already mentioned, the HPSs caused by a daemon in the capture of a nucleus in the ZnS(Ag) layer consist actually of the true HPS component created by the nucleons ejected from the excited nucleus and of the NLS component, which originates from the electrons that surrounded the nucleus initially. Therefore, the daemon-initiated HPSs should lie, on the average, slightly to the left of the HPS peak of the Gaussian in Fig. 2. The Gaussians of each module differ somewhat in reality from one another. They vary with time too. Therefore, because of the fairly moderate statistics obtained in each module, we did not succeed in formulating exact criteria and automating the process of breaking $N(\Delta t)$ down into the $N_w$ and $N_n$ components, with due allowance for the shift of the daemon HPSs to the left wing of the Gaussian. The histograms depicted in Fig. 5 were obtained by selecting one out of several possibilities, subject to the condition that, as stated above, $\int N_w(\Delta t)dt > \int N_n(\Delta t)dt$. As this should be expected in view of *Postulate B,* processing the same data (as well as those collected in April 2002 and 2003) under the simple condition $\int N_w dt \approx \int N_n dt$, reduces slightly the contrast between the $N_w(\Delta t)$ and $N_n(\Delta t)$ distributions; nevertheless, the inverted distribution $N_w(\Delta t) + N_n(-\Delta t)$ can again be fitted with a good statistical confidence by a straight line with $a < 0$. Therefore, the arguments for the existence of the GESCO population remain valid.

A question may naturally arise: if the daemon is capable of traversing many times the Earth or the Sun, how can one conceive of its generating a signal in crossing a 10-μm-thick ZnS(Ag) layer?

From the time taken up by a GESCO daemon to sink along the contracting orbit under the surface of the Earth one readily comes to the estimate that in each passage through the Earth with a velocity ~10 km/s the daemon is slowed down only by $\Delta V \sim 10$ m/s [29]. Indeed, on capturing, say, a Fe nucleus in a ~10 μm layer, it crosses in ~30 μs a distance of ~30 cm while being confined in the remainder of the nucleus it is "digesting". It is primarily over this distance that the daemon is acted upon by the force ~$10^{-10}$ N. Only after the charge of this complex has become negative, the daemon captures another nucleus within a path ~10 μm, a process accompanied by fast (~$10^{-9}$ s) emission of many electrons and nucleons. It is this last process that occurs in our ZnS(Ag) detector layer as the daemon enters it from the air, no part of its kinetic energy being expended in producing the scintillation. The scintillation is excited by the energy released in the capture of the Zn nucleus by the daemon.

**7. Some Consequences of the Detection of Daemons with Respect to the Earth, the Sun, and the Galaxy**

The detection of a daemon population which appears twice a year and disappears under the Earth's surface in two to three months permits us to estimate the number of negative daemons which have built up in the Earth during $4.5 \times 10^9$ years. Accepting the flux $f_\oplus \sim 10^{-9}$ cm$^{-2}$s$^{-1}$ and an average GESCO period of ≈ 6000 s, it amounts to ~$3 \times 10^{23}$. Having a giant mass, daemons form a kernel at the Earth's center [29]. The existence of such a kernel suggests a unique solution for a variety of long-standing problems in geophysics. The most essential of them are: (*i*) the origin of the ~20 TW energy flux escaping from the Earth (the other ~20 TW part can be accounted for by radioactive decay of U, Th, $^{40}$K and gravitational separation of matter); (*ii*) the source of the $^3$He flux; (*iii*) the source to support convection in the lower mantle or even in the iron core, which is needed for generation of the Earth's magnetic field; (*iv*) a source of energy in the lower mantle (or even deeper) to produce about twenty narrow ascending plumes of hot material (Iceland, Hawaii, etc.), which carry also excess $^3$He outwards; (*v*) the reasons for the anisotropy in the seismic properties of the solid(?) iron inner core (IC), where the velocity of the longitudinal waves along the axis of rotation exceeds by 3-4% the velocity of these waves in the azimutal direction; (*vi*) the reasons for azimuthal asymmetry of the IC and possible presence in it of a solid phase. This list could be continued at will (see refs. in Ref. 29).

The fact is that the daemons of the kernel, in capturing the iron nuclei diffusing into it, first excite them, as we have seen, and initiate emission of nucleons and of their clusters ($^2$D, $^3$T, $^3$He, $^4$He, etc.), after which they catalyze proton decay in the remainder of the nucleus with the corresponding liberation of energy. Having specified the total energy release (let it be 10 TW) under the assumption that the external pressure of 360 GPa is balanced by the gas kinetic pressure of daemons on the surface of the self-gravitating isothermal kernel made up of negative and positive daemons, one can readily estimate its parameters by considering the Coulomb diffusion of multiply-charged Fe nuclei into the kernel. It turns out that the size of the kernel is measured in centimeters, and its temperature is ~$10^6$ K, which, in view of the immense mass of the daemons, corresponds to their thermal velocity of only ~50 μm/s. The "digestion" of the iron nuclei and the resultant energy release take place in the near-surface layer of the kernel ~20 Å thick [29].

The energy release in the kernel at the Earth's center provides reasonable answers to the above-mentioned and some other geophysical problems (the anisotropy and asymmetry can be assigned to the effect of the Earth's rotation on the structure of the multiphase thermal convection in the IC; it acquires a two-dimensionl pattern with rolls oriented along the axis of the Earth's rotation), as well as offers a new insight into the history and causes of global tectonics. Because proton decay should in the final count be accompanied by emission of e- and μ-neutrinos with energies <0.1 GeV, the existence of the daemon kernel in the Earth may find support in observation of small maxima in the zenith-angle distributions of sub-GeV μ-like events recorded from the direction to the Earth's center in both Soudan-2 and Super-Kamiokande experiments (see Figs. 29 and 31 in the review of Kajita and Totsuka [30]).

A daemon kernel should exist in the Sun as well [31]. Our optimistic estimates yielded $\sim 2.4 \times 10^{30}$ for the number of daemons in it [17]. If each daemon disintegrated continuously protons at an average rate $(\Delta\tau_{ex})^{-1} \approx 1$ MHz, the released energy would be $\sim 3.6 \times 10^{36}$ erg/s, which practically coincides with the Sun's luminosity. This may be considered as an indication that a noticeable fraction of the energy generated in the Sun is due to the daemon-stimulated proton decay. This is probably argued for by both the well-known deficiency of electron neutrinos in the solar flux and the recently revealed flux of neutrinos of the non-electron flavor [32], in which case there would be no need whatsoever to invoke the concept of neutrino oscillations.

This overall picture meets, however, immediately with difficulties in ensuring involvement of an appreciable fraction of negative daemons in the catalysis of proton decay. If, besides the negative daemons whose charge is practically always compensated by the captured protons or compound nuclei, the kernel contains positive daemons with $Z \sim 10$ as well, the latter for $T > 10^7$ K should be practically completely ionized. For any reasonable assumptions, their large charge should hamper strongly proton diffusion into the kernel (we may recall that in the case of the Earth, where Fe nuclei with $Z_n = 26$ diffuse into the kernel, it is only a thin external layer with a thickness $\sim 10^{-7}$ of the kernel size that accounts primarily for the energy release).

Thus it appears that the idea of the proton decay accounting for a sizable part of the Sun's luminosity may be reasonable if there are negative daemons only. In view of the extremely small number of daemons (there are $\sim 10^{18}$ protons and electrons per daemon in the Universe), this does not come in conflict with anything we know of (does not the barion asymmetry exist?). The repulsive electrostatic action of the kernel daemons is overcome by their gravitation. Moreover, one may conceive here of fascinating cosmological aspects (it is appropriate to recall the hypothesis of Lyttleton and Bondi [33] about the charge asymmetry of the Universe as the cause of its expansion). If the Sun's kernel provides an energy release of $\sim 10^{33}$ erg/s and consists only of $\sim 10^{30}$ negative daemons, its radius should be $\sim 10$ cm for a daemon temperature of $\sim 2 \times 10^{11}$ K and thermal velocity of $\sim 1$ cm/s [31].

It appears natural to believe that the daemon concentration should also grow toward the Galactic center and exceed considerably that in the Galactic halo periphery. The density distribution of the daemons, as in the case of the Earth and the Sun, should approximate to a certain extent the distribution of matter in an isothermal gas sphere. Because, however, of the much rarer (Coulomb) interactions of particles with one another, the prehistory of the ensemble as a whole, i.e., the creation and evolution of galaxies, should play an important role [34].

The γ-radiation emanating from the Galactic center (GC), which had been detected as far back as 1972 in balloon experiments [35] and identified subsequently with the 511-keV positron annihilation line, may be considered as a weighty argument for the daemon hypothesis of the nature of DM. Further observations, including the recent SPI spectrometer measurements on INEGRAL [36], have confirmed this discovery and showed the source of the 511-keV radiation (with a linewidth of only ~3 keV) to be located in the central bulge of the Galaxy and to be practically symmetric, with an FWHM ≈ 9°. No disk component has been detected with certainty. The γ-ray flux is ~$10^{-4}$ ph·cm$^{-2}$·s$^{-1}$, which adds up to yield a total flux from the center of the Galaxy ~$10^{46}$ ph·s$^{-1}$, or ~$2.6\times10^6\,L_\odot$ in the 511-keV line.

The nature of such a strong positron flux has thus far not been revealed unambiguously. One considered as its possible sources neutron stars, black holes, radioactive nuclei like $^{26}$Al, $^{56}$Co etc produced in explosions of supernovas, and so on (for refs. see Refs. 34, 37).

An interesging view is advocated by Boehm *et al* [37], which suggest that the positrons could be generated in mutual annihilation of low-mass (~1-100 MeV) DM particles. The authors point out that in this case the detection of GC positrons may have announced the long-awaited discovery of DM objects. Because the frequency of physical collisions of DM particles with one another is proportional to the square of their concentration, they find that the observed angular distribution of the 511-keV radiation is in a good enough agreement with the expected DM concentration distribution in the GC region.

The daemon-stimulated proton decay entails positron emission. In view of the fact that the concentration of barion matter should be closely correlated with that of the DM objects and, with a good approximation, be proportional to it, so that the frequency of proton capture by daemons should be proportional to the square of the concentration of the latter, it becomes clear that the agreement of the 511-keV radiation distribution with the profile of the central halo DM condensation, when combined with other observations, argues strongly for the daemon hypothesis of the DM nature. Because proton decay should produce, besides positrons, mesons as well, it would be of interest to analyze from the same standpoint the specific features of the γ-radiation in the region ~$10^2$ MeV, which is generated in their decay.

Obviously enough, daemon kernels, which grow continuously with time, should exist and affect noticeably the structure and evolution of the planets, stars and galaxies. These kernels are not full analogs of black holes, because, while converting the surrounding matter to energy, they do not increase in mass. Moreover, this conversion supports their high temperature and pressure, thus preventing the relativistic collapse. One cannot, however, rule out the possibility of such a collapse in the case of much more massive condensations of daemons (quasars, AGNs, etc.).

## 8. General Conclusions and Prospects

The starting point of our work on detection of the DArk Electric Matter Objects, daemons, was a fairly general hypothesis of the existence of relic remnants of the Planckian epoch, likewise Planckian objects, which carry, however, the corresponding electric charge. The scheme chosen for the detection of negative daemons was based on their fairly obvious properties, namely, the possibility of their capturing nuclei with subsequent emission of electrons and nucleons, as well as the daemon-stimulated proton decay. After that, we set the goal of detecting not the low-density DM population of the galactic halo but rather the concentrated low-velocity (~10-30 km/s) population of the Solar system.

The only purpose of our experiments undertaken in the first stage was limited to obtaining any indication of the existence of daemons. We did not have in mind carrying out high-precision quantitative measurements. By now this goal has been reached. In addition to seasonal variations, we have detected many other manifestations of daemons. We have pointed out also a number of consequences following immediately from our experiments, which provide an answer to some long-standing problems.

Despite the apparent simplicity of the detector design, the experiment turned out to be anything but simple. Because of the seasonal variation of the daemon flux, the adjustment and selection of optimum parameters of the detector are determined by the time scale of ~1 year. The properties of the daemon turned out very unconventional for objects of the nuclear physics we got used to. In passing through the detector components, it varies in a certain sense its characteristics by "memorizing" their composition due to the capture and transport of their atomic nuclei and, possibly, molecules and their fragments. The daemon/nuclear-remainder complex with a continuously decreasing $Z_{eff}$ exhibits sequentially the properties of chemically very active elements like Na, F, O, N, C capable of capturing and forming stable molecules with atoms of the substance it traverses. It was found, for instance, that even the black paper employed initially for the mutual light isolation of scintillation screens, as the white writing paper used for reflecting the light, affect the detector properties; indeed, they contain kaolin ($Al_2O_3 \cdot 2SiO_2 \cdot 2H_2O$). There may be other, still unrevealed, factors that are capable of affecting the measurements and reducing their efficiency (see *Postulate B*).

Nevertheless, it may be considered established that: (1) we have discovered low-velocity ($V \sim$ 5-10-30 km/s), highly penetrating objects crossing the detector from both above and below, with (2) the shape of the oscillations produced by them in a thin, ~10 μm, ZnS(Ag) layer depending on the direction of their flight (and, of course, on the conditions of observation), (3) the flux of these objects is $\sim 10^{-9}$ cm$^{-2}$s$^{-1}$, and (4) it varies with a period of 0.5 year.

The available totality of experimental data can be explained in an uncontradictory way in terms of the daemon hypothesis. As a consequence, these data permit estimation of the time of daemon-stimulated proton decay. It was found to be $\Delta\tau_{ex} \sim 1$ μs. Thus, we have apparently observed for the first time manifestation of the phenomena occurring on the Planckian scales ($\sim 10^{19}$ GeV in energy), i.e., quantum gravitation.

One could naturally try to find *a different interpretation of the results obtained in our experiments*.

An interesting possibility has been suggested by Foot and Mitra [38]. They believe that the signals recorded by our detector may be caused by micrometeorites of the mirror matter. The mirror matter constitutes a part of the DM. In interacting with the matter of our Universe, primarily gravitationally, it is capable of traversing large thicknesses of it practically unhindered. Nevertheless, microscopic aggregates of the mirror matter, when moving through an ordinary matter, can develop features of an effective, uncompensated electric charge $\leq 10^{-7}$e, which reduces their range in conventional condensed matter down to ~10 m. Under certain assumptions, such a low-velocity ($V \sim 30$ km/s) particle could, in the opinion of Foot and Mitra, produce a correlated signal in our detector. The half-a-year periodicity of the signals could be due to the crossing by the Earth of the orbit of a mirror micrometeorite flux, and the yearly periodicity observed by DAMA [7], to objects of the galactic halo, whose high velocity ($V \sim 250$ km/s) permits their penetration into the Gran Sasso underground laboratory. The only thing that disagrees radically with the "mirror" interpretation of our experiment is the detection of a particle flux moving upward (the deviation of $N(\Delta t)$ from a constant level for $\Delta t < 0$, see Fig. 5); indeed, mirror micrometeorites cannot traverse the Earth. One cannot, however, rule out the possibility that further development of the theory of mirror matter, made taking into account our data, will eventually remove this obstacle.

Another interesting question is why *no indication of daemon propagation has thus far been observed in other experiments*. One could suggest an answer which allegedly was made by Catherine the Great (the Russian Empress, 1729-1796): "Чего не поищешь, того верно не сыщешь" ("One does not find what he/she is not looking for") [39], but the question should be apparently addressed to the people who carry out the experiments, because who else but they should know the specific features of operation of their detectors best of all.

Nevertheless, without claiming thoroughness of our consideration, we shall try to point out a few obvious and typical reasons for this:
(1) A search for numerous (because of their relatively small mass) WIMPs by detecting the recoil nuclei forces one: (*i*) to use compact detectors (the need of accumulating a large number of kg × day) with a comparatively small area; indeed, with a detector 7 cm in diameter (e.g., the EDELWEISS [5] or CDM-Stanford [4] experiments) and a daemon flux from the halo $\sim 10^{-12}$ cm$^{-2}$s$^{-1}$, one could expect to detect one daemon event in one thousand years(!); (*ii*) to isolate low-energy events (<10 keV), which excludes from consideration the high-energy nuclear events (~0.01-1 GeV) caused by daemons (DAMA [7] being a typical experiment in this respect).
(2) The use of veto systems with a dead time ~1 ms. In this time, a daemon propagating with a velocity of 10 km/s passes a distance of 10 m. The veto system would prevent, for instance, detection of daemon-stimulated proton decays in experiments aimed at detecting the spontaneous proton decay in Super-Kamiokande [40].
(3) Another extreme is looking for superhigh-energy events correlated with cosmic-ray showers. As an illustration may serve the NAUTILUS [41] experiment with a gravitational detector, in which the lower sensitivity threshold was ~3 TeV, and the typical coincidence time interval (and, which is essential, the natural vibration period of the detector) ~1 ms. For the daemon-stimulated proton decay time of ~1 μs, the energy released in this time interval will be only ~1 TeV.
(4) Water and organic scintillators are by far not the best media for daemon detection. The capture of carbon and oxygen nuclei in these materials (as of nitrogen in the air) is hindered by the high excitation potential of their first level (4.4 MeV for $^{12}$C, 6 MeV for $^{16}$O, and 2.3 MeV for $^{14}$N, to be compared with ~1 MeV for Zn).

Thus, in view of the above-mentioned unusual features in the behavior of daemons and in their interaction with matter, seen clearly even in our simple detector, one could hardly hope that standard approaches with the use of sophisticated but operationally limited detecting equipment would result in an accidental discovery of daemons.

*The prospects and directions to be pursued in further research* are pretty obvious:
(1) Manifold (tens of times) increase in the number of scintillation modules aimed at acquisition of good statistics in a few days. This would permit one to establish the effect of variation of the system parameters and introduce the necessary improvements into the detector design in real time (see also item 4 below);
(2) Arrangement of modules one under the other to isolate only daemon-stimulated events by the coincidence of the velocities with which objects traverse each module and the space between them;
(3) Organizing low-background measurements underground to reduce the effect of cosmic rays and natural radioactivity;
(4) A detailed study of the detector properties and of the daemon interaction with matter through variation of the detector parameters (dimensions, the material of the detector components and of additional screens, scintillator type etc). Exploration of the possibilities inherent in non-scintillation detectors;
(5) Designing detectors for observation of daemons from the galactic disk and halo and for the study of the spatial variations in their flux;
(6) Development of the theory of capture of atomic nuclei by the daemon, including the emission of Auger and other electrons and evaporative de-excitation of the nuclei themselves. Consideration of the possibility and consequences of formation and transport of molecules produced in the chemical (valence) capture of atoms and radicals of conventional matter by the remainder of the nucleus, which is gradually "digested" by the daemon contained in it;
(7) A comprehensive study of the physics of the daemon kernels of planets, stars, galaxies etc. and of the consequences of their existence (including neutrino and positron physics);
(8) Experimental and theoretical investigation of the daemon-stimulated proton decay as a necessary basis for construction of a (combined?) (a) theory of quantum gravitation and (b) theory of elementary particles beyond the Standard Model;
(9) Development of various cosmological scenarios with the daemon DM (and the barion matter as a by-product of the daemon ensemble evolution?);
(10) Studying the possibility of practical realization of the daemon-assisted catalysis of the fusion of light nuclei (like carbon) with the aim of constructing a new source of power.

Obviously enough, this list could be continued and broadened in a variety of directions.

**Acknowledgements**

The author is greatly indebted to S. Turck-Chieze for drawing his attention to the problem of the galactic center positrons. Thanks are due also to S.S. Kozlowski, V.M. Moltchanov and M.P. Petrov for valuable discussions which allowed to make the presentation more rigorous.

These results were reported at the International Workshop on Astroparticle and High Energy Physics (AHEP2003, October 13-19, 2003, Valencia, Spain; see http://jhep.sissa.it/cgi-bin/PrHEP/cgi/reader/list.cgi?confid=10).

**References**

1. D.A. Bauer, in: *First Intnl. Workshop on Particle Physics and the Early Universe (COSMO-97)*, L.Roszkowski (ed.), World Scientific (1998), pp.140-154.
2. N.J.C. Spooner, in: *First Intnl. Workshop on Particle Physics and the Early Universe (COSMO-97)*, L.Roszkowski (ed.), World Scientific (1998), pp.155-171.
3. K. Miuchi, M. Minowa, A. Takeda, H. Sekiya, Y. Shimizu, Y. Inoue, W. Ootani and Y. Ootuka, *Astropart. Phys.* **19**, 135 (2003).
4. R. Abusaidi *et al*. (CDMS Collaboration), *Phys. Rev. Lett.* **84**, 5699 (2000).
5. A. Benoit *et al*. (EDELWEISS Collaboration), *Phys. Lett.* **B545**, 43 (2002).
6. B. Ahmed *et al*. (UKDM Collaboration), *hep-ex/0301039* (2003).
7. R. Bernabei *et al*. (DAMA Collaboration), *Riv. Nuovo Cimento* **26**, 1 (2003); *astro-ph/0307403*.
8. J.L. Feng, A.Rajaraman and F. Takayama, *Phys. Rev. Lett.* **91**, 011302 (2003)
9. J.D. Barrow, E.J. Copeland and A.R. Liddle, *Phys. Rev.* **D46**, 645 (1992).
10. S. Alexeyev, A. Barrau, G. Boudoul, O. Khovanskaya and M. Sazhin, *Class. Quantum Grav.* **19**, 4431 (2002).
11. M.A. Markov, *Progr. Theor. Phys.* **Suppl.**, **Extra Number**, 85 (1965); *ZhETF* **51**, 878 (1966).
12. K.P. Stanyukovich, *Doklady Acad. Scis.* **168**, 781 (1966).
13. P. Ivanov, P. Naselsky and I. Novikov, *Phys. Rev.* **D50**, 7173 (1994).
14. H.V. Klapdor-Kleingrothaus and K. Zuber, *Particle Astrophysics,* IOP Publ. Ltd. (1997).
15. A. Linde, private communication (1997).
16. J.H. Bahcall, C. Flynn and A. Gould, *Astrophys. J.* **389**, 234 (1992).
17. E.M. Drobyshevski, *Mon. Not. Royal Astron. Soc.* **282**, 211 (1996).
18. E.M. Drobyshevski, in: *Dark Matter in Astro- and Particle Physics*, H.V.Klapdor-Kleingrothaus and Y.Ramachers (eds.), World Scientific (1997), pp.417-424.
19. E.M. Drobyshevski, On interaction of black miniholes with matter, *Preprint PhTI-1663,* St.Petersburg (1996), pp.1-10.
20. R.N. Cahn and S.L. Glashow, *Science* **213**(4508), 607 (1981).
21. V.A. Rubakov, *Pis'ma v ZhETF* **33**, 658 (1981); *Nucl. Phys.* **B203**, 311 (1982).
22. E.M. Drobyshevski, *Mon. Not. Royal Astron. Soc*. **311**, L1 (2000).
23. E.M. Drobyshevski, *Phys. Atomic Nuclei* **63**, 1037 (2000).
24. E.M. Drobyshevski, *Astron. & Astrophys. Trans.* **21**, 65 (2002).
25. J.B. Birks, *The Theory and Practice of Scintillation Counting,* Pergamon, Oxford (1964).
26. E.M. Drobyshevski, M.V. Beloborodyy, R.O. Kurakin, V.G. Latypov and K.A.Pelepelin, *Astron. & Astrophys. Trans.* **22**, 19 (2003).
27. R.A. Fisher, *Statistical Methods for Research Workers*, 6[th] ed., Oliver & Boyd, Edinburgh, London (1936).
28. E.M. Drobyshevski, M.E. Drobyshevski, T.Yu. Izmodenova and D.S.Telnov, *Astron. & Astrophys. Trans.* **22**, 263 (2003); *astro-ph/*0305597.

29. E.M. Drobyshevski, *Astron. & Astrophys. Trans.* **23**, 49 (2004); *astro-ph*/0111042v2 (2001).
30. T. Kajita and Y. Totsuka, *Revs. Mod. Phys.* **73**, 85 (2001).
31. E.M. Drobyshevski, *Astron. & Astrophys. Trans.* **23** (2004, in press); *astro-ph*/0205353 (2002).
32. Q.R. Ahmad *et al.* (SNO Collaboration), *Phys. Rev.Lett.* **89**, 011301 (2002).
33. R.A. Lyttleton and H. Bondi, *Proc. Roy. Soc.* **A252**, 311 (1959).
34. J.F. Navarro, C.S. Frenk and S.D.M. White, *Ap. J.* **490**, 493 (1997).
35. W.N. Johnson, F.R. Harnden and R.C.Haymes, *Ap. J.* **172**, L1 (1972).
36. J. Knödlseder *et al., Astron. Astrophys.* **407**, L55 (2003); *astro-ph*/0309442v1 (2003).
37. C. Boehm, D. Hooper, J. Silk, M. Casse and J. Paul, *astro-ph*/0309686v3 (2003).
38. R. Foot and S. Mitra, *Phys. Rev.* **D68**, 071901 (2003); *hep-ph*/0306228 (2003).
39. Catherine II, *Choosen Russian Proverbs (Vybornyja Rossijskija Poslovitsy)* (in Russian), Publ. Acad. Scis., St.-Petersburg (1783).
40. M. Shiozawa *et al.* (Super-Kamiokande Collaboration), *Phys. Rev. Lett.* **81**, 3319 (1998).
41. P.Astone *et al.*, *Phys. Lett.* **B499**, 16 (2001).